\documentclass[11pt]{article}
\usepackage{amsmath}
\usepackage{amssymb}
\usepackage{amsthm}
\usepackage[latin1]{inputenc}
\usepackage{graphics,graphicx}

\begin{document}

\title{Vortex solutions of the evolutionary Ginzburg-Landau type equations}
\author{\sc T.Zuyeva\footnote{Permanent address:  Mathematical Division of B.I.Verkin Institute for Low Temperature Physics and Engineering, 
47, Lenin ave., 61103 Kharkov, UKRAINE, e-mail: zueva@ilt.kharkov.ua}}
\date{}


\maketitle

\renewcommand{\baselinestretch}{1.2}
\renewcommand{\theequation}{\thesection .\arabic{equation}}

\centerline{Université Paris-6, Laboratoire Jacques-Louis Lions}

\centerline{175, rue Chevaleret, Paris, 75013, FRANCE\footnote{
e-mail: zuyeva@ann.jussieu.fr}}

\medskip

\begin{abstract} 

We consider two types of the time-dependent Ginzburg-Landau equation in 2D 
bounded domains: the heat-flow equation and the Schr\"{o}dinger equation. We
study the asymptotic behaviour of the vortex solutions of these equations 
when the vortex core size is much smaller than the
inter-vortex distance. 

Using the method of the asymptotic
expansion near the vortices, we obtain the systems of ordinary differential
equations (ODEs) governing
the evolution of the vortices. The expressions for these equations in the
circle and in the annular domain  are presented.
We study the motion of the vortices in these two domains. It is shown that there
exist the stationary points for both types of the equations and  these
points are determined  for some particular cases. 

{\it The heat-flow equation} describes the particles that move like electric 
charges. If the vortices have 
different signs (i.e. Poincaré indices) then 
they attract each other  and collide. If they
have the same signs then they repulse.  The particles always move  away
  from the
stationary points. 
The motion is very slow  and 
 non-periodic. 

{\it The Schr\"{o}dinger equation} describes the 
motion of particles that behave like hydrodynamics vortices. 
The vortices with the same signs move in the same direction. If the
signs are different then the vortices move in the opposite directions. In particular, if the initial positions
are near the stationary  points, then  the particles move along
the elliptic trajectories. 
The motion is not stable with respect to the initial data. It is always periodic or
quasiperiodic. Examples of such trajectories are presented.

\end{abstract}

\newpage
\tableofcontents

\section{Introduction}

We start with the free energy in Ginzburg-Landau theory:
\begin{equation}
\label{eq11}
F(u)=\int_{\Omega}\frac{\hbar^2}{2m}\big\vert \nabla u\big\vert^2+\alpha\vert
u\vert^2+\frac{1}{2}\beta\vert u\vert^4.
\end{equation}

Here in (\ref{eq11}) $u({\bf x},t)$ (${\bf x}=(x,y)$) is a complex-valued wave
function, $\Omega$ is the domain 
occupied by (super)fluid, $m$ is an atomic mass, $2\pi\hbar$ is a Planck's
constant, $\alpha$ and $\beta$ are the temperature-dependent parameters such that  
$$
\alpha(T)=\alpha_0\left(\frac{T}{T_c}-1\right), \quad \beta(T)>0, \quad \alpha_0>0
$$

with $T_c$ being a temperature of phase transition, i.e. $\alpha<0$ for 
$T<T_c$.

In the absence of gradients $F(u)$ represents the 
condensation energy with the density \\
$\displaystyle{f(u)=\alpha\vert u\vert^2+\frac{1}{2}\beta\vert
  u\vert^4}$.

We assume that there are $N$ vortices in the domain, located at the points
${\bf \xi}_1(t)$,\ldots,${\bf \xi}_N(t)$, ${\bf \xi}_j=(\zeta_j,\eta_j)$ with degrees $n_1$,\ldots,$n_N$, respectively. The vortices
are stable if and only if $n_j=±1$. If 
$\varepsilon$ is a size of vortex core, than inter-vortex distance is supposed to be of
the order  $O(1)$. 

Let us pass to dimensionless variables 
$u=u_0u^\prime$, ${\bf x}=\varepsilon {\bf x}^{\prime}$, where $\displaystyle{u_0^2=-\frac{\alpha}{\beta}}$
corresponds to the minimum of the condensation energy and
$\displaystyle{\varepsilon=\sqrt{\frac{\hbar^2}{2m\vert \alpha\vert}}}$ is a coherence
length (size of vortex core). Then the free energy takes the form
\begin{equation}
\label{eq12}
F(u^{\prime})=\int_{\Omega}\vert\nabla^\prime u^\prime\vert^2+\frac{1}{2}\left(1-\vert u^{\prime}
  \vert^2\right)^2
\end{equation}

Note that in these variables the vortex core size is $O(1)$ and
inter-vortex distance is $O(1/\varepsilon)$.

Using  standard techniques of the calculus of
variation, the minimization of $F$ with respect to variation in $u^\prime$ yields the
stationary G-L equation:
\begin{equation}
\label{eq13}
\Delta u^\prime +u^\prime(1-\vert u^\prime\vert^2)=0.
\end{equation}

In this paper we study two types of non-stationary Ginzburg-Landau equations:
the nonlinear Schr\"{o}dinger equation
\begin{equation}
\label{eq151}
\Delta u^\prime +u^\prime(1-\vert u^\prime\vert^2)=-iu^{\prime}_t;
\end{equation}
and the nonlinear heat flow equation:
\begin{equation}
\label{eq161}
\Delta u^\prime +u^\prime(1-\vert u^\prime\vert^2)=u^{\prime}_t.
\end{equation}

Under the usual diffusive scaling ${\bf x}^{\prime\prime}={\bf x}^{'}\varepsilon$,
$t^{\prime\prime}=\varepsilon^2t$, Eq. (\ref{eq151}) takes the form
\begin{equation}
\label{eq15}
\Delta u^{\varepsilon} +\frac{1}{\varepsilon^2} u^{\varepsilon}(1-\vert u^{\varepsilon}\vert^2)=-iu^{\varepsilon}_t
\end{equation}
and for (\ref{eq161}) we obtain
\begin{equation}
\label{eq16}
\Delta u^{\varepsilon} +\frac{1}{\varepsilon^2} u^{\varepsilon}(1-\vert u^{\varepsilon}\vert^2)=u^{\varepsilon}_t
\end{equation}
 
We have again the vortex core size $O(\varepsilon)$ and inter-vortex distance $O(1)$.

In this paper we consider these equations in the bounded domain
with the {\it Neumann boundary condition} on $\partial \Omega$.

We look for two approximate solutions of (\ref{eq15}) and  (\ref{eq16}) (see \cite{E}):
{\it outer solution} 
\begin{equation}
v^{\varepsilon}({\bf x},t)=u_0({\bf x},t)+\varepsilon u_1({\bf x},t)+\varepsilon^2u_2({\bf x},t)+\ldots
\label{eq17} 
\end{equation}
in the region outside the vortex cores, but for small $r=\vert{\bf
  x}-{\bf\xi}_j(t)\vert$, and {\it inner solution} 
for the core of the $j$th vortex:
\begin{equation}
U^{\varepsilon}({\bf x},t)=U_0\left(\frac{{\bf x}-{\bf \xi}_j(t)}{\varepsilon}\right)+\varepsilon
U_1\left(\frac{{\bf x}-{\bf\xi}_j(t)}{\varepsilon}\right)+\varepsilon^2U_2\left(\frac{{\bf
    x}-{\bf \xi}_j(t)}{\varepsilon}\right)+\ldots
\label{eq18} 
\end{equation}
Substituting the expansions (\ref{eq17}) and
(\ref{eq18}) in the equations (\ref{eq15}) and
(\ref{eq16}), we
obtain the series of equations for the functions $u_0({\bf x},t)$, $u_1({\bf
  x},t)$, $u_2({\bf x},t)$,\ldots, and $U_0({\bf x},t)$, $U_1({\bf x},t)$,\ldots. We
construct the approximate form for the solution (\ref{eq17}) {\it near
  the $j$th vortex}. For the 
inner solution (\ref{eq18}) we pass to the stretched variable
$\displaystyle{{\bf X}=\frac{{\bf x}-{\bf\xi}_j(t)}{\varepsilon}}$ and find the
solution for large $R=\vert {\bf X}\vert$. Asymptotic matching of these
solutions yields the equations governing the motion of the particles. 
Taking into account that $r=\varepsilon R$,
we obtain that the vortices
described by {\it the Schr\"{o}dinger equation} are gouverned by the system
\begin{equation}
\label{eq19} 
{\bf\dot \xi}_j(t)=2{\bf K}^{(j)},
\end{equation}
where the vector ${\bf K}^{(j)}$ depends on the form of the domain $\Omega$ and can be
expressed, for example, through the derivative of the complex potential $W(z)$ of the fluid with
vortices out of $j$th vortex:
\begin{equation} 
\label{eq110}
{\bf K^{(j)}}=(Re \tilde W^{\prime}(z_j), -Im \tilde W^{\prime}(z_j)).
\end{equation}

In particular, for annular domain we have (see formula (\ref{25}) below):
$$
\tilde W^{\prime}(z_j)=\frac{1}{z_j}\sum_{k=1\atop k\neq j}^N n_k\left[\zeta\left(i\ln\frac{z_j}{z_k}\right)-
\zeta\left(i\ln\frac{z_jz_k}{Z^2_k}\right)
\right]-\frac{n_j}{z_j}\zeta\left(2i\ln\frac{z_j}{R_2}\right)+\frac{2i\eta}{\omega_1z_j}\sum_{k=1}^N n_k\ln\frac{r_k}{R_2}.
$$

We obtain by the same matching procedure for {\it the heat-flow equation} that for
the standard  time scaling $t^{\prime\prime}=\varepsilon^2t$ the vortices are stationary at the
leading order. To obtain a 
non-trivial dynamics we have to rescale the time by the factor
$(\log 1 / \varepsilon)$. In this long-time scaling the dynamics of the vortices
is described by the system
\begin{equation}
\label{eq112} 
{\bf \dot \xi}_j^{\perp}(t)=2n_j{\bf K}^{(j)},
\end{equation}
where $\dot \xi_j^{\perp}=(-\dot \eta_j,\dot \zeta_j)$ and ${\bf K}^{(j)}$ is
determined by 
(\ref{eq110}).

We construct the numerical solutions of the systems (\ref{eq19}) and (\ref{eq112})
in the circle and in the ring for given initial positions of vortices and given indexes. We
show that the Schr\"{o}dinger equation (\ref{eq19}) discribes the motion of
the hydrodynamical vortices and the heat-flow equation (\ref{eq112}) governes
the motion of some charged-similar  particles. The trajectories of the particles for
different initial data are shown in the Figs.1-9.


\section{Schr\"{o}dinger equation}
\setcounter{equation}{0}     

Let us consider the Schr\"{o}dinger equation (\ref{eq15}). Substituting (\ref{eq17})
into (\ref{eq15}), we obtain the following the series of equations:
\begin{equation}
\label{eq201}
\varepsilon^{-2}: \quad u_0(1-\vert u_0\vert^2)=0;
\end{equation}
\begin{equation}
\label{eq202}
\varepsilon^{-1}: \quad u_1\overline{u_0}+u_0\overline{u_1}=0;
\end{equation}
\begin{equation}
\label{eq203}
\varepsilon^{0}: \quad -iu_{0t}=\Delta u_0-\vert u_1\vert^2u_0.
\end{equation}

Here $\overline{u}$ means the complex conjugation of $u$.
The first of these equations yields
\begin{equation}  
\label{21}
u_0({\bf x},t)=e^{i\Phi_0({\bf x},t)}.
\end{equation}  
Now from (\ref{eq203}) and   (\ref{eq201}) we get 
$$
\Phi_{0t}=i\Delta\Phi_0-(\nabla\Phi_0)^2-\vert u_1\vert^2.
$$
Separating real and imaginary part, we obtain
$$
\Delta\Phi_0=0.
$$

Asymptotic boundary conditions at the cores are 
$$
\Phi_0({\bf x},t)\to n_j\tan^{-1}\left(\frac{y-\eta_j}{x-\zeta_j}\right)= n_j\theta
$$ 
as
${\bf x}\to{\bf \xi}_j$, 
where $(r,\theta)$ are   the
polar coordinates related to the $j$th vortex: $r=\vert {\bf x-\xi}_j\vert$, 
$\theta=\tan^{-1}\displaystyle{\left(\frac{y-\eta_j}{x-\zeta_j}\right)}$.

Hence, a boundary-value problem for $\Phi_0({\bf x},t)$ takes the form:
\begin{equation}  
\label{22}
\begin{cases}
\Delta\Phi_0({\bf x},t)=0,  \hskip 3cm \rm{in}\quad \Omega\setminus\{\cup_{j=1}^N {\bf \xi}_j\}; \cr\cr
\Phi_0({\bf x},t)\to n_j\theta,\hskip 2.5cm {\bf x}\to{\bf \xi}_j, \quad j=1,2,\ldots,N\cr\cr
\displaystyle{\frac{\partial\Phi_0}{\partial n}=0} \hskip 4cm {\bf x}\in\partial\Omega. \cr
\end{cases}
\end{equation}

The last boundary condition follows from the condition for $u^{\varepsilon}$
on $\partial\Omega$. It is well-known (see e.g. \cite{LL}) that the phase of condensate 
wave function $\Phi({\bf x},t)$ is related the hydrodynamic velocity potentilal
$\phi({\bf x},t)$ 
$$
\phi({\bf x},t)=\frac{\hbar}{m}\Phi({\bf x},t).
$$

Then 
we obtain  ${\bf v}=\nabla\phi=\displaystyle{\frac{\hbar}{m}\nabla \Phi}$, and the boundary condition is the standard hydrodynamics condition of impermeability, ${\bf
  v}\cdot{\bf n}=0$. Thus we can consider problem (\ref{22}) as a usual
hydrodynamic problem describing the potential of the ideal liquid with the vortices.

The solution of this problem can be found for any form of the domain $\Omega$ by the method of ``reflexed
vortices'' or by some other method. In
particular, for the annular domain$R_1\leq r\leq R_2$ the velocity potential
$\Phi_0({\bf x},t)$ of an ideal fluid with 
$N$ vortices of vortex strengths $n_k$ can be written in the complex form
(see. \cite{Z1})
$$
\Phi_0(z)=Re W(z)=Re\left\{-i\sum_{k=1}^N
  n_k\left[\ln \sigma\left(i\ln\frac{z}{z_k}\right)-
\right.\right.
$$
\begin{equation}
\label{23}
\left.\left.
-\ln \sigma\left(i\ln\frac{zz_k}{Z^2_k}\right)-\frac{2\eta}{\omega_1}\ln\frac{r_k}{R_2}\ln z\right]\right\}+const.
\end{equation}

Here $z$ is a complex variable: $z=x+iy=r\exp(i\tilde \theta)$,
$z_k=\zeta_k+i\eta_k=r_k\exp(i\tilde \theta_k)$ are the points of vortex locations, 
$r=\vert{\bf x}\vert$,
the polar angle 
$\tilde \theta=\tan^{-1}\left(\frac{y}{x}\right)$ is measured {\it from the center
of the ring},  $W(z)$ is a complex
potential, $Z_k=R_2\exp(i\theta_k)$, $\zeta(z)$ and  $\sigma(z)$ are the
Weirstrass zeta- and sigma-function with the half-periods
$\omega_1=\pi$, $\omega_2=i\ln(R_2/R_1)$; $\eta=\zeta(\omega_1)$.(see e.g. \cite{AS}).
 
To obtain the approximate expression for the wave function $u_0$ {\it near the $j$th
vortex}, let us expand $\Phi_0$ in a Taylor series around $j$th vortex. We have to take into
account that $\Phi_0$ has singularities $n_j\tan^{-1}(y-\eta_j)/(x-\zeta_j)$ at the points of
vortex locations ${\bf \xi}_j$. Extracting this singularitity, i.e. representing  $\Phi_0({\bf x},t)=n_j\theta+\Phi_{0j}({\bf x},t)$
and expanding $\Phi_{0j}({\bf x},t)$ in the series
\begin{equation} 
\label{241}
\Phi_{0j}({\bf x},t)=\Phi_{0j}({\bf \xi}_j,t)+\nabla\Phi_{0j}({\bf \xi}_j,t)\cdot({\bf x}-{\bf \xi}_j)+\ldots,
\end{equation}
we obtain the relation
\begin{equation} 
\label{24}
u_0({\bf x},t)=e^{i\Phi_0}\approx e^{in_j\theta+i\theta_0}[1+i{\bf K}^{(j)}\cdot({\bf x-\xi}_{\j})+O(r^2)],
\end{equation}
where
$$
\theta_0=\Phi_{0j}({\bf \xi}_j,t)
$$
and
\begin{equation} 
\label{251}
{\bf K^{(j)}}=\nabla\Phi_{0j}({\bf \xi}_j,t)=(Re \tilde W^{\prime}(z_j), -Im \tilde
W^{\prime}(z_j)).
\end{equation}
Here $\tilde W^{\prime}(z_j)$ is a derivative of the complex potential in the point of the
vortex location. 

In particular, for annular domain we have (see \cite{Z2})
\begin{equation} 
\label{25}
\tilde W^{\prime}(z_j)=\frac{1}{z_j}\sum_{k=1\atop k\neq j}^N n_k\left[\zeta\left(i\ln\frac{z_j}{z_k}\right)-
\zeta\left(i\ln\frac{z_jz_k}{Z^2_k}\right)
\right]-\frac{n_j}{z_j}\zeta\left(2i\ln\frac{r_j}{R_2}\right)+\frac{2i\eta}{\omega_1z_j}\sum_{k=1}^N n_k\ln\frac{r_k}{R_2}.
\end{equation}

One can show that the expression (\ref{25}) gives
the velocity potential of the ideal liquid with the vortices {\it it the
  circle} when $R_1\to 0$ (see Section 5)
\begin{equation} 
\label{252}
\tilde W^{\prime}(z_j)=\sum_{k=1\atop k\neq j}^N \frac{n_k}{i(z_j-z_k)}- \sum_{k=1}^N
\frac{n_k}{i(z_j-z^{\prime}_k)},
\end{equation}
where $z^{\prime}_k=\displaystyle{\frac{R_2^2}{z_k} }$ are the
coordinates of the vortices symmetric to $z_k$ with respect to the
circumference $r=R_2$.

Thus, the exteriour solution  $u_0({\bf x},t)$ is defined by (\ref{24}),
(\ref{251}) and (\ref{25}) (or (\ref{252})).

Let us consider now the core structure of the vortices. We denote by 
${\bf X}=\displaystyle{\frac{{\bf x}-{\bf \xi}_j(t)}{\varepsilon}}$ the stretched
variable and by $(R,\theta)$ are the polar coordinates of ${\bf X}$. Then the inner solution (\ref{eq18})
takes a form
$$
U^{\varepsilon}({\bf x},t)=U_0({\bf X})+\varepsilon U_1({\bf X})+\varepsilon^2U_2({\bf X})+\ldots
$$

Substituting this expansion in the equation (\ref{eq16}) we obtain the series of 
equations for $U_0({\bf X})$, $U_1({\bf X})$, $U_2({\bf X})$\ldots:
\begin{equation}
\Delta U_0+U_0(1-\vert U_0\vert^2)=0,
\label{26}
\end{equation}
\begin{equation}
\Delta U_1+(1-2\vert U_0\vert^2)U_1-U_0^2\overline{U_1}=i\dot {\bf \xi}_j(t)\cdot\nabla U_0,
\label{27}
\end{equation}

As in the consideration above (see.  (\ref{24})) we
look for the solution of (\ref{26}) in the form:
$$
U_0(R,\theta,t)=f_0(R)e^{in_j\theta+i\theta_0}.
$$
Then $f_0(R)$ satisfies the equation
$$
f_0^{\prime\prime}+\frac{1}{R}f_0^{\prime}-\frac{n_j^2}{R^2}f_0+f_0(1-f_0^2)=0
$$
with boundary conditions
$$
\begin{cases}
f_0(0)=0, \cr
f_0(\infty)=1. \cr
\end{cases}
$$

As in \cite{E} it can be shown that 
\begin{equation}
f_0(R)\approx1-\frac{n_j^2}{2R^2}+ O\left(\frac{1}{R^4}\right)
\label{28}
\end{equation}
for  $R\gg 1$. 

As for $U_1({\bf X})=U_1(R,\theta)$, we look for it in the form
\begin{equation}
U_1(R,\theta)=f_1(R,\theta)e^{in_j\theta+i\theta_0}=[(A_r(R)+iA_i(R))\cos\theta+(B_r(R)+iB_i(R))\sin\theta]
e^{in_j\theta+i\theta_0}.
\label{29}
\end{equation}

Substituting $U_0$ and $U_1$ in (\ref{27}), gathering  the terms with
$\sin\theta$ and $\cos\theta$ and separating real and imaginary parts, we obtain the
system of equations for $A_r$, $A_i$, $B_r$ and $B_i$: 
$$
-n_j\dot\eta_j\frac{f_0}{R}=A_r^{\prime\prime}+\frac{1}{R}A_r^{\prime}+\left(1-3f_0^2-\frac{2}{R^2}\right)A_r-
n_j\frac{2B_i}{R^2},
$$
$$
n_j\dot\zeta_j\frac{f_0}{R}=B_r^{\prime\prime}+\frac{1}{R}B_r^{\prime}+\left(1-3f_0^2-\frac{2}{R^2}\right)B_r+n_j\frac{2A_i}{R^2},
$$
$$
f_0^{\prime}\dot\zeta_j=A_i^{\prime\prime}+\frac{1}{R}A_i^{\prime}+\left(1-f_0^2-\frac{2}{R^2}\right)A_i+n_j\frac{2B_r}{R^2},
$$
$$
f_0^{\prime}\dot\eta_j=B_i^{\prime\prime}+\frac{1}{R}B_i^{\prime}+\left(1-f_0^2-\frac{2}{R^2}\right)B_i-n_j\frac{2A_r}{R^2}.
$$

We look for the solutions $A_r$, $A_i$, $B_r$, and $B_i$ in the form
$$
A_r=-n_j\dot\eta_jW(R), \quad A_i=\dot\zeta_j Z(R),
$$
\begin{equation}
B_r=n_j\dot\zeta_j W(R), \quad B_i=\dot\eta_j Z(R).
\label{291}
\end{equation}

Substituting these expressions into equations for  $A_r$-$B_i$ we derive two
equations for $W(R)$ and $Z(R)$:
\begin{equation}
\begin{cases}
\displaystyle{\frac{f_0}{R}=W^{\prime\prime}+\frac{1}{R}W^{\prime}+W\left(1-3f_0^2-\frac{2}{R^2}\right)+\frac{2}{R^2}Z},\cr\cr
\displaystyle{f_0^{\prime}=Z^{\prime\prime}+\frac{1}{R}Z^{\prime}+W\left(1-f_0^2-\frac{2}{R^2}\right)+\frac{2}{R^2}W}.\cr
\end{cases}
\label{292}
\end{equation}

Taking into account (\ref{28}), we conclude that
$$
f_0^{\prime}= \frac{1}{R^3}+ O\left(\frac{1}{R^5} \right)
$$
for large $R$.

Expanding $W(R)$ and $Z(R)$ in the series with respect to the degrees of $R$,
we can find the solution $U_1(R,\theta,t)$  {\it for large R}:
\begin{equation}
\label{210}
U_1(R,\theta,t)=e^{in_j\theta+i\theta_0}\left[\frac{n_jC_1}{R}\left(
    {\bf\dot\xi^{\perp}}_j\cdot\frac{\bf
        X}{R}\right)+\frac{iR}{2}\left({\bf\dot\xi}_j\cdot\frac{\bf
          X}{R}\right)+\frac{iA_1}{R}\left(
    {\bf\dot\xi}_j\cdot\frac{\bf X}{R}\right)+ O\left(\frac{1}{R^3}\right)\left({\bf\dot\xi}_j\cdot\frac{\bf X}{R}\right)\right],
\end{equation}
where ${\bf \dot\xi}_j=({\bf \dot\zeta}_j,{\bf \dot\eta}_j)$, ${\bf
 \dot \xi^{\perp}}_j=(-\dot \eta_j,\dot \zeta_j)$, $C_1$, $A_1$ are constants. 

Then the  solution $U({\bf X})=U_0({\bf X})+\varepsilon U_1({\bf X})$ can be
represented as
\begin{equation}
\label{211}
U({\bf X})=e^{in_j\theta+i\theta_0}\left[1+\frac{i\varepsilon}{2}({\bf \dot \xi}_j\cdot{\bf X})+
  O\left(\frac{1}{R^2}\right)\cdot\left(1+\varepsilon \left({\bf \dot \xi}^{\perp}_j\cdot\frac{\bf X}{R}\right)\right)\right]
\end{equation}
for large $R$.

The function $U({\bf X})$ is an approximate solution in the region near the $j$th vortex,
where $R=\vert({\bf x-\xi}_j)\vert/\varepsilon$ is large.

To guarantie consistency of the solutions (\ref{24}) and (\ref{211}) 
it is necessary to suppose that the leading terms of the expansion be equal for 
${\bf x-\xi}_j(t)=\varepsilon{\bf X}$, i.e.
$$
{\bf K}^{(j)}\cdot\varepsilon{\bf
  X}=\frac{1}{2}{\bf\dot\xi}_j\cdot\varepsilon{\bf X}.
$$
This gives the equations
\begin{equation}
\label{212}
{\bf\dot\xi}_j=2{\bf K}^{(j)}, \qquad j=1,2,\ldots,N.
\end{equation}

This is the desired system
describing the motion of the vortices in the domain $\Omega$.


\section{Heat flow equation}
\setcounter{equation}{0}

Now we consider the heat-flow equation (\ref{eq16}). To construct the solution
near the $j$th vortex in outer region we substitute (\ref{eq17}) in
(\ref{eq16}) and obtain the series of
equations for $u_0$, $u_1$,\ldots:
$$
u_0(1-\vert u_0\vert^2)=0;
$$
$$
u_1\overline{u_0}+u_0\overline{u_1}=0;
$$
$$
u_{0t}=\Delta u_0-\vert u_1\vert^2u_0.
$$

For the function $u_0$ we have
\begin{equation}  
\label{31}
u_0({\bf x},t)=e^{i\Phi_0({\bf x},t)},
\end{equation}  
as it was in the case of the the Scr\"{o}dinger equation, 
but now $\Phi_0$ satisfies the equation:
\begin{equation}  
\label{32}
\Phi_{0t}=\Delta\Phi_0.
\end{equation}

Passing to the {\it moving} coordinates $\tilde {\bf x}={\bf x}-{\bf \xi}_j(t)$ we obtain  the
following equation for $\Phi_0$ that describes the local behaviour of the solution
near the $j$th vortex:
\begin{equation}  
\label{33}
\begin{cases}
\Phi_{0t}-{\bf\dot \xi}_j(t)\cdot\nabla\Phi_0=\Delta\Phi_0({\bf \tilde x},t),   \cr\cr
\Phi_0({\bf \tilde x},t)\to n_j\tan^{-1}\left(\tilde y/ \tilde x\right)=n_j\theta,\hskip 1.5cm
{\bf \tilde x}\to0; \cr
\end{cases}
\end{equation}

We denote by $(r,\theta)$ the polar coordinates corresponding to ${\bf\tilde
  x}$. The function  $\Phi_0$ {\it
  near the vortex} is described by the power series of $r$
and $\log r$ for small $r$. It can be shown that
\begin{equation}  
\label{34}
\Phi_0(\tilde {\bf x},t)=n_j\theta+\frac{1}{2}n_j(\log r)\dot {\bf \xi}_j^{\perp}\cdot 
\tilde {\bf x}+{\bf S_j\cdot \tilde x}+ O(r^2\log r),
\end{equation}
where ${\bf S}_j$ is an unknown vector. 
For the function $u_0$ we have
\begin{equation}  
\label{35}
u_0(\tilde {\bf x},t)=e^{n_j\theta}[1+\frac{i}{2}n_j(\log r)\dot {\bf \xi}_j^{\perp}\cdot {\bf \tilde x}+i{\bf S}_j\cdot {\bf\tilde x}+ O(r^2\log r)].
\end{equation}

This gives a representation of the outer solution near the $j$th vortex.

The inner solution for the heat flow equation in the stretched variables leads
to the system of equations for $U_0$, $U_1$,\ldots, similar to (\ref{26})-(\ref{27}):
\begin{equation}
\Delta U_0+U_0(1-\vert U_0\vert^2)=0,
\label{36}
\end{equation}
\begin{equation}
-\dot {\bf \xi}_j(t)\cdot\nabla U_0=\Delta U_1+(1-2\vert U_0\vert^2)U_1-U_0^2\overline{U_1}.
\label{37}
\end{equation}

The first equation is the same as (\ref{26}) and hence we have again
(\ref{28}). We look for the function  $U_1({\bf X})$ in the form
(\ref{29}). Then 
Eq.(\ref{37}) in terms of $f_0$, $f_1$ becomes
\begin{equation}
-{\bf \dot\xi}_j\cdot(f_0^{\prime}\nabla R+in_j f_0\nabla \theta)=\Delta f_1+2in_j(\nabla
f_1\cdot\nabla\theta)-\frac{n_j^2}{R^2}f_1+f_1(1-2f_0^2)-f_0^2\overline f_1.
\label{38}
\end{equation}

The following system for $A_r(R)$, $A_i(R)$, $B_r(R)$, $B_i(R)$ is valid:
$$
-\dot\zeta_jf_0^{\prime}=A_r^{\prime\prime}+\frac{1}{R}A_r^{\prime}+\left(1-3f_0^2-\frac{2}{R^2}\right)A_r-
n_j\frac{2B_i}{R^2},
$$
$$
-\dot\eta_jf_0^{\prime}=B_r^{\prime\prime}+\frac{1}{R}B_r^{\prime}+\left(1-3f_0^2-\frac{2}{R^2}\right)B_r+n_j\frac{2A_i}{R^2},
$$
$$
-n_j\dot\zeta_j\frac{f_0}{R}=A_i^{\prime\prime}+\frac{1}{R}A_i^{\prime}+\left(1-f_0^2-\frac{2}{R^2}\right)A_i+n_j\frac{2B_r}{R^2},
$$
$$
n_j\dot\eta_j\frac{f_0}{R}=B_i^{\prime\prime}+\frac{1}{R}B_i^{\prime}+\left(1-f_0^2-\frac{2}{R^2}\right)B_i-n_j\frac{2A_r}{R^2}.
$$

We seek the solution of this system in the form that is similar to (\ref{291}):
$$
A_r=\dot\zeta_j Z, \quad A_i=\dot\eta_j n_j W, \quad B_r=\dot\eta_j Z, \quad B_i=-\dot\zeta_j n_j W, 
$$
where $Z$ and $W$ satisfy
$$
-f_0^{\prime}=Z^{\prime\prime}+\frac{1}{R}Z^{\prime}+\left(1-3f_0^2-\frac{2}{R^2}\right)Z+\frac{2W}{R^2},
$$ 
$$
-\frac{f_0}{R}=W^{\prime\prime}+\frac{1}{R}W^{\prime}+\left(1-f_0^2-\frac{2}{R^2}\right)W+\frac{2Z}{R^2}.
$$

The solutions $W$ and $Z$ can be found in the power series of $R$
and $\log R$. We are interesting in the behaviour of the solutions for $R\gg 1$:
$$
W=-\frac{1}{2}R\log R+C_0R+ O(\log R),
$$
$$
Z=-\frac{1}{2R}\log R+ O\left(\frac{1}{R}\right).
$$

Here $C_0$ is a constant. So, we have for  $R\gg 1$
\begin{equation} 
U_1({\bf X})=e^{in_j\theta}\left[\frac{1}{2}in_j(\log R)(\dot\xi_j^{\perp}\cdot {\bf X})+
in_j C_0(\dot\xi_j^{\perp}\cdot {\bf X})+ O\left(\frac{\log R}{R}\right)\right].
\label{39}
\end{equation}

Putting together the solutions $U_0$ and
$U_1$ we obtain as $R\to \infty$
\begin{equation} 
U_0({\bf X})+\varepsilon U_1({\bf X})=e^{in_j\theta}\left[f_0(R)+
\frac{1}{2}i\varepsilon n_j(\log R)(\dot\xi_j^{\perp}\cdot {\bf X})+
i\varepsilon n_j C_0(\dot\xi_j^{\perp}\cdot {\bf X})+ O\left(\varepsilon\frac{\log R}{R}\right)\right].
\label{310}
\end{equation}

Matching (\ref{310}) with the outer solution (\ref{35}) and taking into account
that ${\bf \tilde x}=\varepsilon{\bf X}$, $r=\varepsilon R$, we can see that
$$
\frac{1}{2}(\log\varepsilon){\bf\dot\xi}_j^{\perp}+n_j{\bf S}_j=
C_0{\bf\dot\xi}_j^{\perp}
$$
or
\begin{equation} 
{\bf\dot\xi}_j^{\perp}=\frac{{\bf S}_j}{-\frac{1}{2}\log\varepsilon+C_0}= O(1/\log(1/\varepsilon)).
\label{311}
\end{equation}

This equation shows that in this time scale the vortices are {\it stationary} at the
leading order. The time scale of the vortices is slower than the time scale of 
the phase field.


\section{Heat flow equation: long time scale}
\setcounter{equation}{0}

To obtain a non-trivial dynamics of the vortices, we have to rescale the time
variable by a factor $O(\log(1/\varepsilon))$, i.e. instead of diffusive
scaling ${\bf x}^{\prime\prime}={\bf x}^{'}\varepsilon$, $t^{\prime\prime}=\varepsilon^2t$ we consider the 
scaling ${\bf x}^{\prime\prime}={\bf x}^{'}\varepsilon$,
$t^{\prime\prime}=[\varepsilon^2/\log(1/\varepsilon)]t$. Then we obtain  the equation (see \cite{E})
\begin{equation}
\label{41}
\Delta u^{\varepsilon} +\frac{1}{\varepsilon^2} u^{\varepsilon}(1-\vert u^{\varepsilon}\vert^2)=\frac{1}{\log(1/\varepsilon)} u^{\varepsilon}_t.
\end{equation}

We consider the approximate solutions of this equation in the form
(\ref{eq17}) for outer region and (\ref{eq18}) for inner region. Thus we obtain
as in the previous case the equations (\ref{21}) for $u_0$. For the function $\Phi_0$ we have
\begin{equation}
\label{42}
\frac{1}{\log(1/\varepsilon)}\Phi_{0t}=\Delta\Phi_0.
\end{equation}

To investigate the behaviour of the outer solution near the $j$th vortex we expand
$\Phi_0$ in powers of $\delta=(\log(1/\varepsilon))^{-1}$:
$$
\Phi_0=\Phi_{00}({\bf x},t)+\delta\Phi_{01}({\bf x},t)+\delta^2\Phi_{02}({\bf x},t) +\ldots
$$

Then we get the series of equations for $\Phi_{0i}$:
\begin{equation}
\label{43}
\Delta\Phi_{00}=0,
\end{equation}
\begin{equation}
\label{44}
\Phi_{00,t}=\Delta\Phi_{01}, \ldots
\end{equation}

Eq.(\ref{43}) shoud be completed by the boundary conditions at the vortices: 
$$
\Phi_{00}({\bf x},t)\to n_j\tan^{-1}\left(\frac{y-\eta_j}{x-\zeta_j}\right)
$$
and the Neumann condition on the boundary $\partial\Omega$, i.e. $\displaystyle{\frac{\partial\Phi_{00}}{\partial n}=0}$.

So, we have problem (\ref{22})  for $\Phi_{00}({\bf x},t)$ and hence
$\Phi_{00}({\bf x},t)$ is given by (\ref{23}). Thus, near the $j$th vortex 
$$
\Phi_{00}({\bf x},t)=n_j\theta+\theta_0+{\bf K}^{(j)}\cdot({\bf x-\xi}_j(t)),
$$
where ${\bf K^{(j)}}$ is defined by (\ref{25}).

For the function $\Phi_{01}$
we have 
$$
\Phi_{01}({\bf x},t)=\frac{1}{2}n_j(\log r)({\bf\dot \xi}_j^{\perp}\cdot 
\tilde {\bf x})+{\bf P}_{1j}\cdot{\bf\tilde x}+ O(r^2\log r)
$$
for $r=\vert\tilde {\bf x}\vert\ll 1$, 
${\bf P}_{1j}$ is a constant vector depending on the history of the
vortices. Extracting the singularitity of $\Phi_{01}({\bf x},t)$ 
in the $j$th vortex (see (\ref{241})),
we obtain 
$$
u_0({\bf \tilde x},t)=e^{in_j\theta}\left(1+i{\bf K}^{(j)}\cdot {\bf\tilde x}+
\frac{in_j}{2\log(1/\varepsilon)}(\log r)({\bf \xi}_j^{\perp}\cdot 
\tilde {\bf x})+\frac{in_j}{\log(1/\varepsilon)}(\log r)({\bf P}_{1j} \cdot 
\tilde {\bf x})\right)+
$$
\begin{equation}
\label{45}
+ O(\delta^2+\delta r^2\log r+r^2)
\end{equation}
as $r$ is sufficiently small.

It follows from (\ref{41}) that for {\it the inner solution} 
\begin{equation}
\label{46}
-\frac{\dot {\bf \xi}_j(t)}{\log(1/\varepsilon)}\cdot\nabla U_0=\Delta U_1+(1-2\vert U_0\vert^2)U_1-U_0^2\overline{U_1}.
\end{equation}

By the same way as it was done for (\ref{37}) we obtain
\begin{equation}
U_0({\bf X})+\varepsilon U_1({\bf X})=
e^{in_j\theta}\left[f_0(R)+
\frac{in_j}{2\log(1/\varepsilon)}(\log R)(\dot\xi_j^{\perp}\cdot {\bf X})+
\frac{in_j C_0}{\log(1/\varepsilon)}
(\dot\xi_j^{\perp}\cdot {\bf X})+ O\left(\frac{\varepsilon\log R}{R
\log(1/\varepsilon)}\right)\right]
\label{47}
\end{equation} 
as $R\to\infty$.

Matching (\ref{45}) and (\ref{47}) gives 
\begin{equation}
\label{48}
{\bf K}^{(j)}+\frac{n_j}{2\log(1/\varepsilon)}(\log
\varepsilon)\dot\xi_j^{\perp}+\frac{n_j}{\log(1/\varepsilon)}{\bf P}_{1j}=
\frac{n_j C_0}{\log(1/\varepsilon)}
\dot\xi_j^{\perp}.
\end{equation} 

Therefore we obtain the desired equation
$$
\dot\xi_j^{\perp}=
2n_j{\bf K}^{(j)}+O(1/\log(1/\varepsilon)).
$$ 

In this time scale the dynamics of vortices is not trivial and it is
described by the system of equations
\begin{equation}
\label{49}
\dot\xi_j^{\perp}=2n_j{\bf K}^{(j)}, \hskip 1cm j=1,2,...,N.
\end{equation} 

This equation is similar to (\ref{212}) for the Schr\"{o}dinger equation, but it
includes the turned coordinates  $\dot \xi_j^{\perp}$ and the factor $n_j$. Notice
that the time scale in (\ref{49}) is slower by the factor 
$O(\log(1/\varepsilon))$.


\section{Motion of the vortices in the circle}
\setcounter{equation}{0}

Let us show that from (\ref{23}) we obtain the velocity
of the liquid with the vortices in the {\it circle}: 
\begin{equation} 
\label{61}
\tilde W^{\prime}(z)=\sum_{k=1}^N \frac{n_k}{i(z-z_k)}- \sum_{k=1}^N
\frac{n_k}{i(z-z^{\prime}_k)},
\end{equation}
as $R_1\to 0$.
Here $z^{\prime}_k=\displaystyle{\frac{R_2^2}{\overline z_k} }$ are the
coordinates of the vortices symmetric to $z_k$ with respect to the
circumference $r=R_2$.

It follows from (\ref{23}) that 
\begin{equation} 
\label{62}
W^{\prime}(z)=\frac{1}{z}\sum_{k=1}^N n_k\left[\zeta\left(i\ln\frac{z}{z_k}\right)-
\zeta\left(i\ln\frac{zz_k}{Z^2_k}\right)
\right]-\frac{n_j}{z}\zeta\left(2i\ln\frac{r}{R_2}\right)+\frac{2i\eta}{\omega_1z}\sum_{k=1}^N n_k\ln\frac{r_k}{R_2}.
\end{equation}

Notice that as $R_1\to 0$ the denominator $q$ tends to zero. Therefore, the  $\zeta$-function
$$
\zeta(z)=\frac{\eta z}{\omega_1}+ \frac{\pi}{2\omega_1}\left[\cot \frac{\pi z}{2\omega_1}+
4\sum_{t=1}^{\infty}\frac {q^{2t}}{1-q^{2t}}\sin \left(\frac{\pi t z}{\omega_1}\right)\right]
$$
is reduced in the sum of the linear part and cotangents. Hence 
$$
W^{\prime}(z)=\sum_{k=1}^N \frac{n_k}{2z}\left[\cot\left(
\frac{i}{2}\ln\frac{z}{z_k}\right)-
\cot\left(\frac{i}{2}\ln\frac{zz_k}{Z^2_k}\right)\right].
$$

Calculating all the cotangents, we obtain the expression
$$
W^{\prime}(z)=\frac{i}{2}\sum_{k=1}^N n_k
\left[\frac{z+z_k}{z(z-z_k)}-\frac{zz_k+Z^2_k}{z(zz_k-Z^2_k)}\right].
$$

Denoting by $z_k{^\prime}=Z^2_k/z_k$ the coordinates of the vortex symmetric with the
vortex $z_k$ with respect to the circumference $r=R_2$ ($Z_k=R_2\exp(\theta_k)$) and expanding the
terms into the simplest fractions we derive the desired formula 
(\ref{61}). 

To obtain the velocity of the $j$th vortex we have to substract  the
term corresponding to this vortex and then put $z=z_j$. We have 
\begin{equation} 
\label{63}
\tilde W^{\prime}(z_j)=\sum_{k=1\atop k\neq j}^N \frac{n_k}{i(z_j-z_k)}- \sum_{k=1}^N
\frac{n_k}{i(z_j-z^{\prime}_k)}.
\end{equation}

Let us investigate the motion of the vortices in the circle. The systems
(\ref{212}) and (\ref{49}) describing this motion can be represented as 
\begin{equation}
\label{51}
\begin{cases}
\dot\zeta_j=2 K_{j1}, \cr
\dot\eta_j=-2 K_{j2}, \qquad j=1,2,\ldots,N
\end{cases}
\end{equation} 
for  the Scr\"{o}dinger equation and as  
\begin{equation}
\label{52}
\begin{cases}
\dot\zeta_j=-2n_j K_{j2}, \cr
\dot\eta_j=-2n_j K_{j1}, \qquad j=1,2,\ldots,N
\end{cases}
\end{equation} 
for the heat-flow equation.

Here ${\bf K}_j=(K_{j1}, K_{j2})=(\Re W^{\prime}(z_j),\Im W^{\prime}(z_j))$ and
$W^{\prime}(z_j)$ is defined by (\ref{63}).


\subsection{Schrödinger equation}

Notice that the particles described by the Schrödinger equation behave as
hydrodynamic vortices. So, we can expect that they move along the closed
trajectories and never meet the walls. 

If at the initial moment we have the symmetric locations of the vortices with
alternating signs, we obtain the symmetric closed trajectories (see. Fig.1 for
$N=2$ and $N=4$). The closer are vortices to the wall at $t=0$ the wider are trajectories. 

\bigskip

\begin{figure}[h]
\unitlength1cm
\begin{minipage}[t]{7.0cm}
\includegraphics[width=6.0cm]{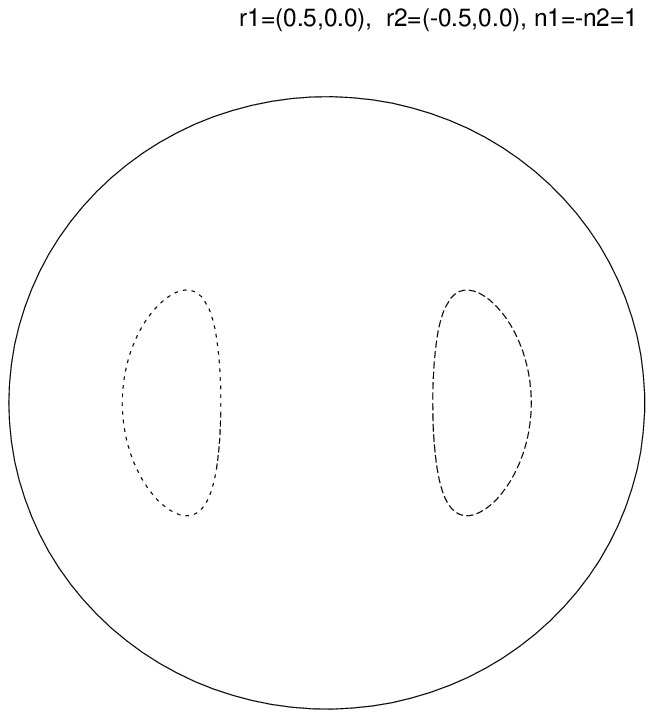}
\begin{center} Fig.1a. Motion ``vortex-antivortex''. $N=2$.
\end{center}

\label{fig1}
\end{minipage}
\hskip 2cm
\begin{minipage}[t]{7.0cm}
\includegraphics[width=6.0cm]{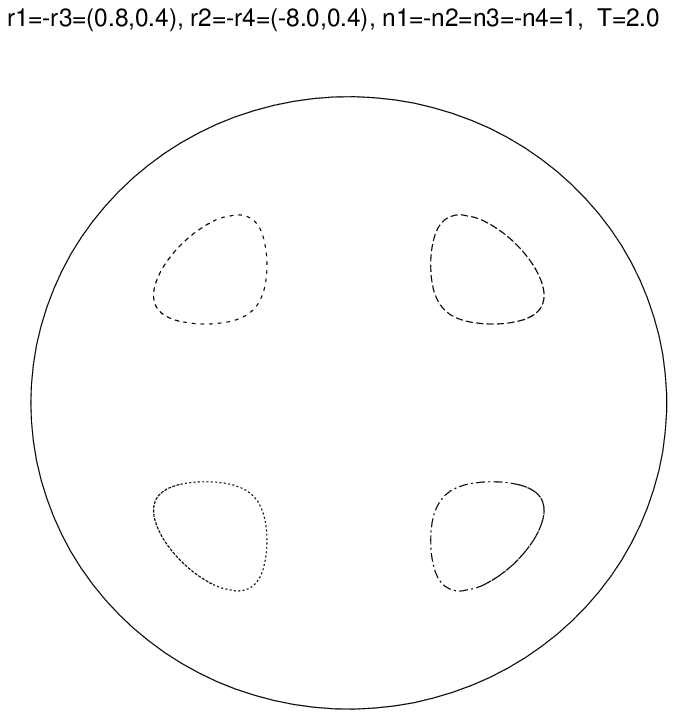}
\label{fig2}
\begin{center} Fig.1b. Motion "vortices-antivortices". $N=4$. 
\end{center}
\end{minipage}
\end{figure}
 
If one of the vortices is close to the center and the other is near the wall we have
more complicated motion: the vortex near the wall moves almost along the circumference 
and the inner vortex passes a ``star'' trajectory (Fig.2a). But if all the
vortices have the same signs and are on the same circumference at $t=0$, they
move independently along this circumference (Fig.2b).

\bigskip
\begin{figure}[h]
\unitlength1cm
\begin{minipage}[t]{7.0cm}
\includegraphics[width=6.0cm]{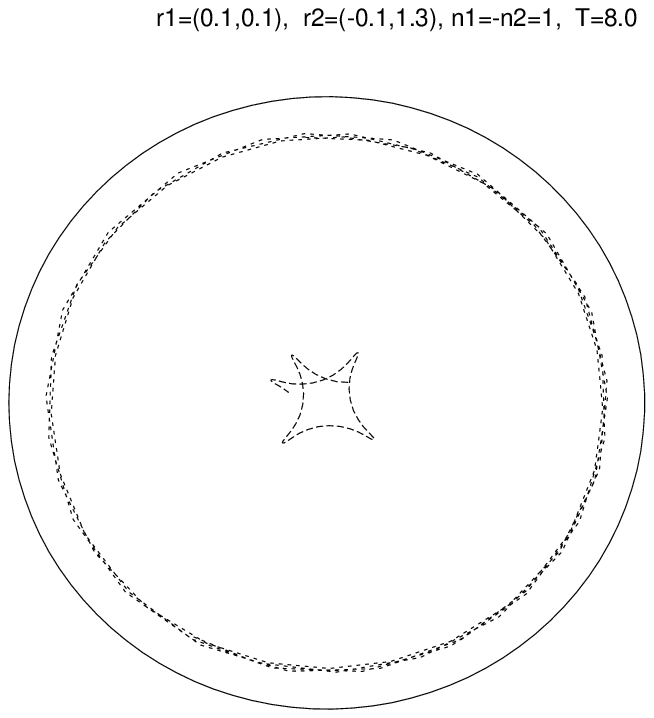}
\begin{center} Fig.2a. Motion ``vortex-antivortex''. Non-symmetric initialy positions.
\end{center}

\label{fig1a}
\end{minipage}
\hskip 2cm
\begin{minipage}[t]{7.0cm}
\includegraphics[width=6.0cm]{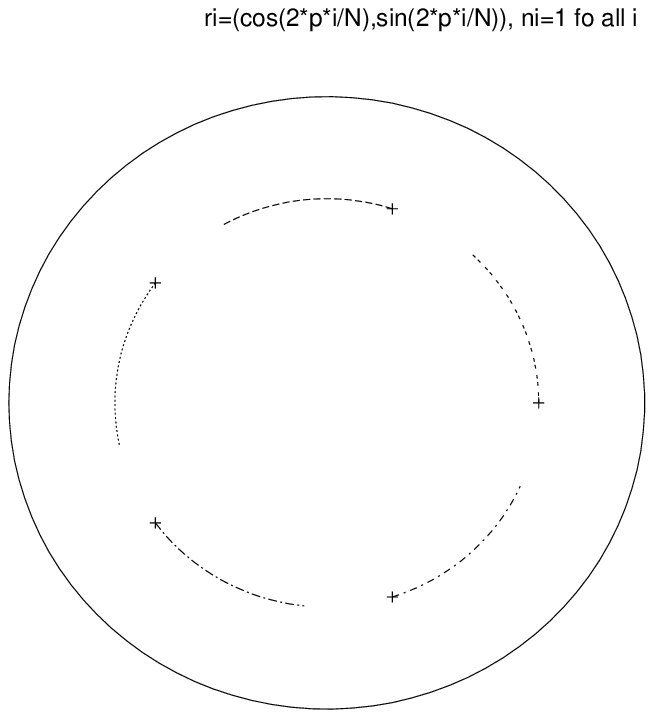}

\label{fig2b}
\begin{center} Fig.2b. Motion of the same vortices on the same circumference. $N=5$.
\end{center}
\end{minipage}
\end{figure}
 
The motion is much more complicated 
for odd number of the vortices with the different signes  (Fig.3a, 3b):
vortices move in the 
chaotic way even if the initial positions are on the same
circumference (comp. with Fig.2b).

\bigskip

\begin{figure}[h]
\unitlength1cm
\begin{minipage}[t]{7.0cm}
\includegraphics[width=6.0cm]{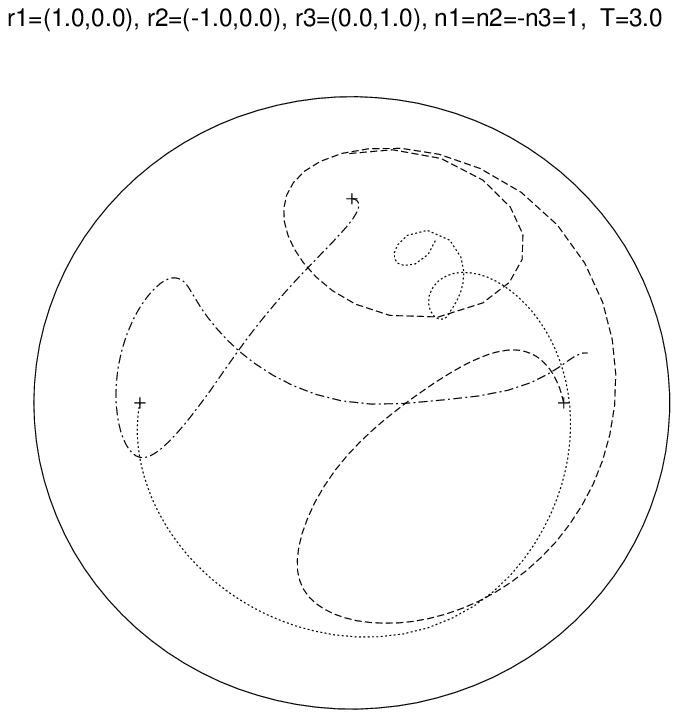}
\begin{center} Fig.3a. Motion ``vortices-antivortices''. Symmetric initialy positions.
\end{center}

\label{fig1a}
\end{minipage}
\hskip 2cm
\begin{minipage}[t]{7.0cm}
\includegraphics[width=6.0cm]{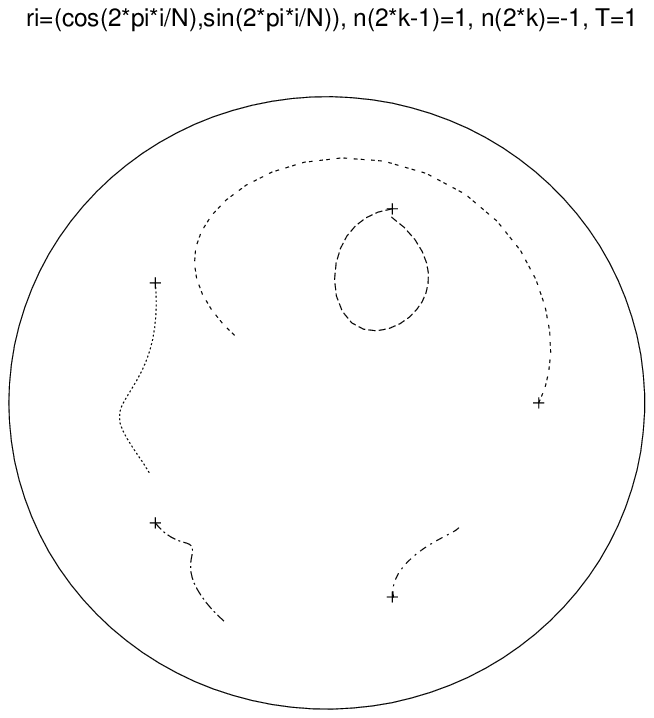}

\label{fig2b}
\begin{center} Fig.3b. $N=5$. 
\end{center}
\end{minipage}
\end{figure}


\subsection{Heat flow equation}

The figures below represent the motion of the particles governed
by the heat flow equation.
In fact, we have only two types of motions: attraction, if
the vortices have different signs and in the initial time are close enough to 
each other, and repulsion, in all other cases.

So, if we have the even number of particles with alternative signs, each
particle is 
attracted  to the nearest neighbour and annihilated (Fig.4a, 4b). 

\bigskip
\begin{figure}[h]
\unitlength1cm
\begin{minipage}[t]{7.0cm}
\includegraphics[width=6.0cm]{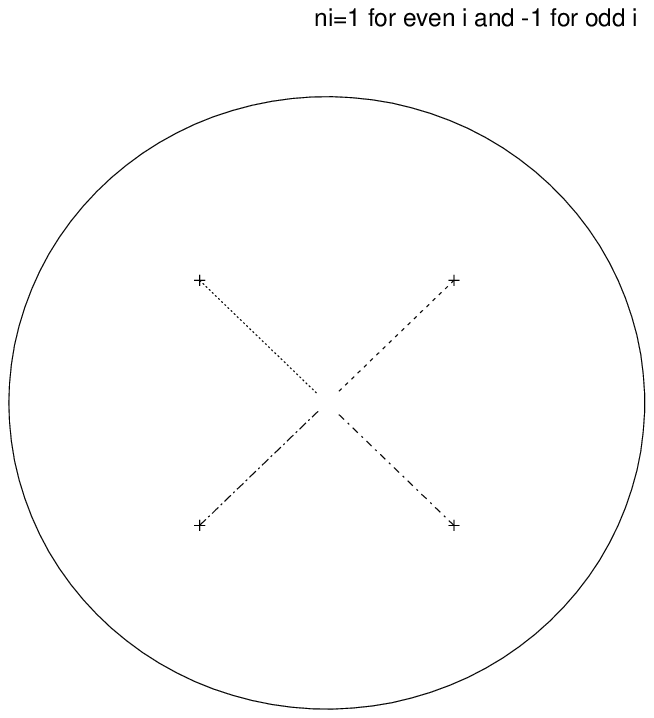}
\begin{center} Fig.4a. Motion of different-sign particles: symmetric initial positions
\end{center}

\label{fig1a}
\end{minipage}
\hskip 2cm
\begin{minipage}[t]{7.0cm}
\includegraphics[width=6.0cm]{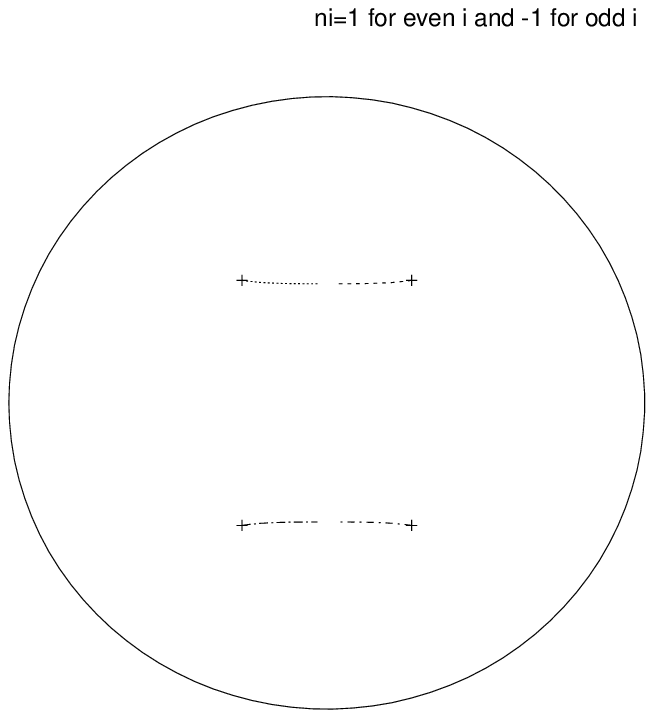}

\label{fig2b}
\begin{center} Fig.4b. Attraction to the nearest neihbour.
\end{center}
\end{minipage}
\end{figure}

For the odd number of particles the picture is symmetric if the initial data is
symmetric. In Fig.5a we have the circle of particles with alternative
signs. In this case the nearest particles with different signs are attracted and
annihilated. 
 
\begin{figure}[h]
\unitlength1cm
\begin{minipage}[t]{7.0cm}
\includegraphics[width=6.0cm]{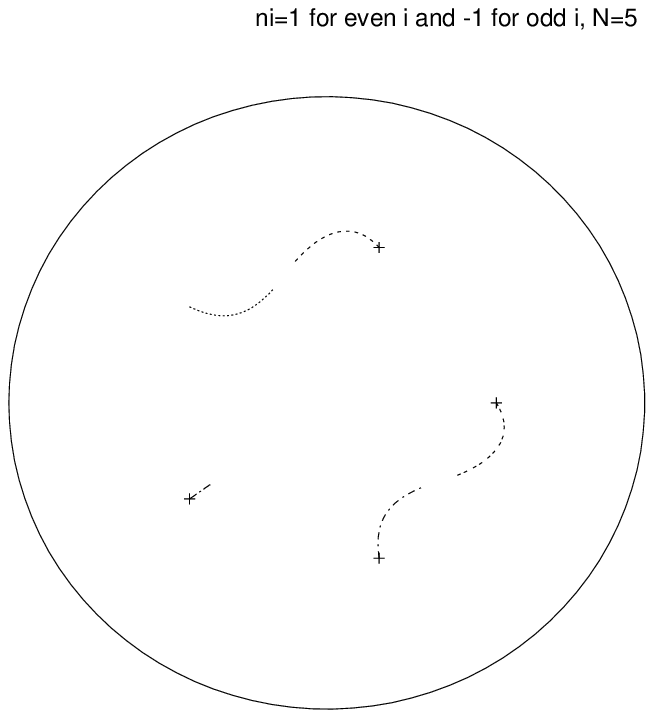}
\begin{center} Fig.5a. Motion of odd number of different-sign particles.
\end{center}

\label{fig1a}
\end{minipage}
\hskip 2cm
\begin{minipage}[t]{7.0cm}
\includegraphics[width=6.0cm]{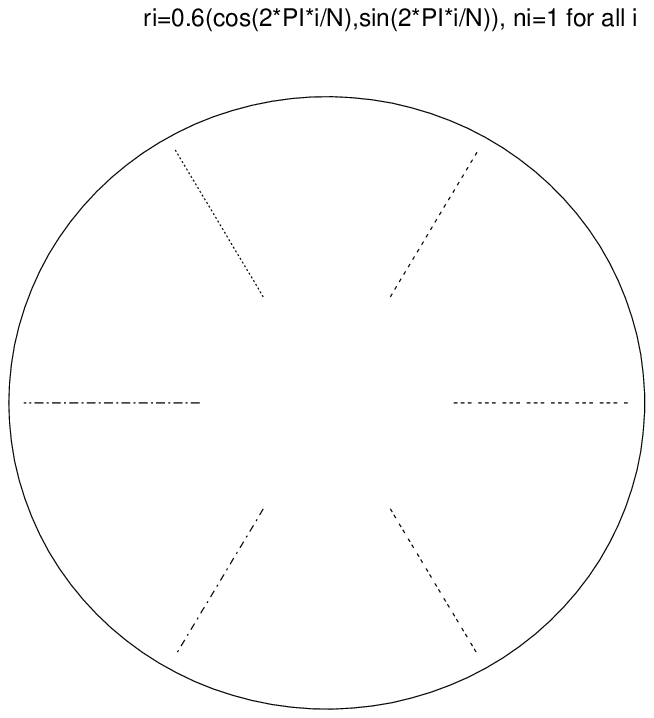}

\label{fig2b}
\begin{center} Fig.5b. Attraction to the boundary.
\end{center}
\end{minipage}
\end{figure}

Finally, any number of particles with the same signs is attracted to the
boundary and annihilated with the imaginary vortices (Fig. 5b).

Thus the motion of the particles in the framework of the heat flow equation is
always attractive or repulsive. There are no periodic and quasiperiodic
trajectories.

\newpage

\section{Motion of the particles in the ring}
\setcounter{equation}{0}

\subsection{Stationary points}

We investigate the motion of the vortices in the {\it ring}. The
motion is governed by   (\ref{51}) and (\ref{52}) but 
$W^{\prime}(z_j)$ is defined by (\ref{25}).

Consider the expression for $W^{\prime}(z_j)$.

\medskip

{\bf A.} Let us suppose that there are only two vortices:  $N=2$. Representing 
$z_j=r_je^{i\theta_j}$ we obtain 
$$
W^{\prime}(z_j)=\frac{1}{z_j}\left[n_k
\zeta\left(i\ln\frac{r_j}{r_k}-(\theta_j-\theta_k)\right)-n_k
\zeta\left(i\ln\frac{r_kr_j}{R_2^2}-(\theta_j-\theta_k)\right)-
\right.
$$
\begin{equation}
\left.
-n_j\zeta\left(2i\ln\frac{r_j}{R_2}\right)+
\frac{2i\eta}{\omega_1}\left(n_1\ln\frac{r_1}{R_2}+
n_2\ln\frac{r_2}{R_2}\right)\right]\equiv \frac{1}{z_j}F_j.
\label{W}
\end{equation}

Here $j,k=1,2, k\neq j$ (if $j=1$, then $k=2$ and vice versa, if  $j=2$, then
$k=1$). 

Notice that $\omega_1=\pi$ and $\omega_2=i\ln(R_2/R_1)$ are the half-periods of the 
$\zeta$-functions. 

If we put $r_1=r_2=r_0$ (vortices are on the same circumference) and take into
account that $\zeta(-z)=-\zeta(z)$), we obtain
$$ 
W^{\prime}(z_j)=\frac{1}{z_j}\left[n_k \zeta\left(\theta_k-\theta_j\right)-
n_k \zeta\left(2i\ln\frac{r_0}{R_2}-(\theta_j-\theta_k)\right)-
n_j\zeta\left(2i\ln\frac{r_0}{R_2}\right)+\frac{2i\eta}{\omega_1}(n_1+n_2)\ln\frac{r_0}{R_2}\right].
$$

Using the addition formula for the $\zeta$-function (\cite{AS}, p.635) we obtain
$$
\zeta\left(2i\ln\frac{r_0}{R_2}-(\theta_j-\theta_k)\right)=
\zeta\left(2i\ln\frac{r_0}{R_2}\right)+
\zeta\left(\theta_k-\theta_j\right)+\frac{1}{2}\cdot\frac{{\cal
    P}^{\prime}\left(\displaystyle{2i\ln\frac{r_0}{R_2}}\right)
    -{\cal P}^{\prime}\left(\theta_k-\theta_j\right)}{{\cal
    P}\left(\displaystyle{2i\ln\frac{r_0}{R_2}}\right)
    -{\cal P}\left(\theta_k-\theta_j\right)},
$$
where ${\cal P}(z)$ is Weierstrass' ${\cal P}$-function (\cite{AS}). Thus 
$$ 
W^{\prime}(z_j)=\frac{1}{z_j}\left[-(n_k+n_j)\left[
  \zeta\left(2i\ln\frac{r_0}{R_2}\right)-
  \frac{2i\eta}{\omega_1}\ln\frac{r_0}{R_2}\right]-
  \frac{n_k}{2}\cdot\frac{{\cal
    P}^{\prime}\left(\displaystyle{2i\ln\frac{r_0}{R_2}}\right)
    -{\cal P}^{\prime}\left(\theta_k-\theta_j\right)}{{\cal
    P}\left(\displaystyle{2i\ln\frac{r_0}{R_2}}\right)
    -{\cal P}\left(\theta_k-\theta_j\right)}\right].
$$

Notice that the function  ${\cal P}^{\prime}(±\omega_i)=0$ vanicshes  at the points of
the half-periods $±\omega_i$, $i=1,2$. Thus, if we put \\
$\displaystyle{2i\ln\frac{r_0}{R_2}=-\omega_2=-i\ln\frac{R_2}{R_1}}$ and 
$\theta_k-\theta_j=\omega_1=\pi$ the   last term will be equal zero. Hence, for
$r_1=r_2=\sqrt{R_1R_2}\quad$, $\theta_j-\theta_k=\pi$  we have
\begin{equation}
\label{53}
W^{\prime}(z_j)=\frac{n_1+n_2}{z_j}\left[\eta^{\prime}-\frac{\eta\omega_2}{\omega_1}\right]=
-\frac{n_1+n_2}{z_j}\cdot\frac{\pi i}{2\omega_1}=
\begin{cases}
0, \qquad n_1=-n_2, \cr
\displaystyle{-\frac{i}{z_j}}, \qquad n_1=n_2=1.
\end{cases}
\end{equation} 

Here we have used the Legendre's relation $\eta\omega_2-\eta^{\prime}\omega_1=\pi i/2$;
$\eta^{\prime}=\zeta(\omega_2)$ (\cite{AS}).

Expression (\ref{53}) means that there exist the stationary points
$r_1=r_2=\sqrt{R_1R_2}\quad$, $\theta_1-\theta_2=\pi$ for both equations if
$n_1=-n_2$. Without loss of generality we can suppose that $n_1=-n_2=1$ and
that the stationary points are 
\begin{equation}
\label{54}
(x_1^0,y_1^0)=(\sqrt{R_1R_2},0), \qquad (x_2^0,y_2^0)=(-\sqrt{R_1R_2},0).
\end{equation}

If we take the initial positions of the vortices at these points, the
vortices will not move at all.

\medskip
{\bf B.} Let us obtain the similar points for any {\it even} number of
vortices $2N$. It was shown  (\cite{Z1})
that if we have $N$ vortices with the equal intensities $n_j=n$ located uniformly {\it on the same circumference $r=r_0$},
i.e. $z_j=r\exp(i\theta_j)$, $\theta_j=2\pi j/N$, $j=1,2,\ldots,N$, the expression for
$W^{\prime}(z)$ take the form
\begin{equation}
\label{541}
W^{\prime}(z)=\frac{n_j}{z}\left[\zeta\left(i\ln\frac{z}{z_j}\right)-
\zeta\left(i\ln\frac{zz_j}{R_2^2}\right)-\frac{2i\eta}{\omega_1}\ln\frac{r}{R_2}\right],
\end{equation}
where $z_j=r_0\exp(i\theta_j)$ is the position of any vortex and the
half-periods of the $\zeta$-function are $\omega_1=\pi/N$ and
$\omega_2=i\ln(R_2/R_1)$. This case corresponds to the {\it vortex chaine} in 
the ring.

If we have $N$ pairs ``vortex-antivortex'' located uniformly on the
circumference, one can interprete this as two vortex chaines. 
The complex-conjugated velocity has the form 
\begin{equation}
\label{542}
W^{\prime}(z)=\sum_{k=1}^2\frac{n_k}{z}\left[\zeta\left(i\ln\frac{z}{z_k}\right)-
\zeta\left(i\ln\frac{zz_k}{R_2^2}\right)-\frac{2i\eta}{\omega_1}\ln\frac{r}{R_2}\right],
\end{equation}
where 
$z_1$ (respectively $z_2$) is the coordinate of a vortex (respectively antivortex) in the
chaines and $n_1=1$, $n_2=-1$.  For the sake of simlicity we can take
$z_1=r_0$ and 
$z_2=r_0\exp(\pi i/ N)$.

To obtain the velocity of the $j$th vortex, we should
substract the term corresponding to this vortex and then put $z=z_j$. We have 
\begin{equation}
\label{543}
W^{\prime}(z_j)=\frac{n_k}{z_j}\left[\zeta\left(i\ln\frac{z_j}{z_k}\right)-
\zeta\left(i\ln\frac{z_jz_k}{R_2^2}\right)-\zeta\left(2i\ln\frac{z_j}{R_2}\right)
\frac{2i\eta}{\omega_1}\ln\frac{r_j}{R_2}\right].
\end{equation}
where if $j=1$, then $k=2$ and if  $j=2$, then $k=1$.

Notice that the real semi-period of the $\zeta$-functions is equal to   
$\omega_1=\pi/N$ while the imaginary semi-period does not change, i.e. 
$\omega_2=i\ln(R_2/R_1)$. As in the part {\bf A} we obtain
that 
\begin{equation}
\label{544}
W^{\prime}(z_j)=\frac{n_1+n_2}{z_j}\left[\eta^{\prime}-\frac{\eta\omega_2}{\omega_1}\right]=
-\frac{n_1+n_2}{z_j}\cdot\frac{\pi i}{2\omega_1}=
\begin{cases}
0, \qquad n_1=-n_2, \cr
\displaystyle{-\frac{iN}{z_j}}, \qquad n_1=n_2=1
\end{cases}
\end{equation} 
for $r_1=r_2=\sqrt{R_1R_2}\quad$, $\theta_j-\theta_k=\pi/N$.  
This  means that the stationary points for two chains of alternating-sign
particles are
\begin{equation}
\label{545}
\begin{cases}  
z_k=r_0 e^{i\pi k/N}, \qquad r_0=\sqrt{R_1 R_2}, \qquad k=1,2,\ldots,2N,\cr
n_k=(-1)^{k+1}.
\end{cases}
\end{equation} 

It occurs that there exist the stationary points in the case when all the vortices have the same signs $n_1=n_2=\ldots=n_N=1$
and are on the 
same circumference $r=r_0$. To find these points we have to solve the equation 
$W^{\prime}(z_j)=0$, where $W^{\prime}(z_j)$ is the velocity of $j$ vortex obtained 
from (\ref{541}) by substraction of the term corresponding to the vortex,
\begin{equation}
\label{546}
W^{\prime}(z_j)=\frac{n}{z_j}\left[-\zeta\left(2i\ln\frac{z_j}{R_2}\right)+\frac{2i\eta}{\omega_1}\ln\frac{r_0}{R_2}\right].
\end{equation}
  
We have
$$
\coth
\left(N\ln\frac{r_0}{R_2}\right)=4\sum_{t=1}^{\infty}\frac{q^{2t}}{1-q^{2t}} 
\sinh\left(2tN\ln\frac{r_0}{R_2}\right),
$$
where $q=\displaystyle{\exp\left(-\frac{\pi\vert \omega_2\vert}{\omega_1} \right)}$ is
the denominator of the $\zeta$-functions.

This equation has a root for any $N$ and
any $R_1$,
$R_2$. For example, if we take  $R_1=0.5$, $R_2=1.5$ and $N=2$ (these parametres were
used in the numerical simulations below) then $r_0=0.651159$. This value
coinsides with the numerically calculated stationary point of the systems (\ref{51}), (\ref{52}).


\subsection{Motion "vortex-antivortex": linear problems}

Let us investigate the motion of the pair ``vortex-antivortex" in the ring,
i.e. we have $n_{1}=-n_{2}=1$ and the stationary points are defined by (\ref{54}).  

If the initial data are {\it
  close to  the stationary points} one can describe  the motion of two
vortices in an analitic way. 
Taking into account that the stationary points
correspond to the solution of the equation $W^{\prime}(z_j)=0$ we obtain
$$
W^{\prime}(z_j)\approx \frac{1}{z_j}\left[\frac{\partial F_j}{\partial x_1}(x_1-x_1^0)+\frac{\partial F_j}{\partial x_2}(x_2-x_2^0)+
\frac{\partial F_j}{\partial y_1}(y_1-y_1^0)+\frac{\partial F_j}{\partial y_2}(y_2-y_2^0)\right],
$$
where $F_j$ is defined in (\ref{W}),
$(x_1^0,y_1^0)$, $(x_2^0,y_2^0)$ are the stationary points, 
$Z_k=R_2\exp(i\theta_k)$, and 
$$
\frac{\partial F_j}{\partial x_j}=\frac{ix_j}{R_1 R_2}\left(n_k(e_2-e_1)+2n_je_3+\frac{2\eta 
    n_j}{\pi}\right); \hskip 2cm
\frac{\partial F_j}{\partial x_k}=\frac{x_k
  n_k}{R_1 R_2}\left(e_1+e_2+\frac{2\eta}{\pi}\right), \quad k\neq j;
$$
$$
\frac{\partial F_j}{\partial y_j}=-\frac{x_jn_k}{R_1 R_2}(e_2-e_1); \hskip 3cm
\frac{\partial F_j}{\partial y_k}=\frac{x_kn_k}{R_1 R_2}(e_2-e_1).
$$

Here we have taken into account that $z_1=x_1^0=r_0=\sqrt{R_1 R_2}$,
$z_1=x_2^0=-r_0=-\sqrt{R_1 R_2}$,  $\zeta^{\prime}(z)=-{\cal P}(z)$ \cite{AS} and 
$$
{\cal P}\left(i\ln\frac{r_kr_j}{R_2^2}-(\theta_j-\theta_k)\right)={\cal
  P}(\omega_1+\omega_2)=e_2;
$$
$$
{\cal P}\left(i\ln\frac{r_j}{r_k}-(\theta_j-\theta_k)\right)={\cal
  P}(\omega_1)=e_1; \qquad {\cal P}\left(2i\ln\frac{r_j}{R_2}\right)={\cal
  P}(\omega_2)=e_3
$$
at the stationary points $(x_1^0,y_1^0)$, $(x_2^0,y_2^0)$. Here  
$e_1$, $e_2$, $e_3$ are the roots of the characteristic equation
(\cite{AS}) 
and $e_1+e_2+e_3=0$. Denoting
$$
a_1=\frac{2}{R_1R_2}\left(e_1-e_2\right),
$$
$$
a_2=\frac{2}{R_1R_2}\left(-e_3+\frac{2\eta}{\pi}\right),
$$
$$
a_3=\frac{2}{R_1R_2}\left(e_2-e_1-2e_3-\frac{2\eta}{\pi}\right),
$$
we can represent $W^{\prime}(z_j)$ in the form
$$
W^{\prime}(z_1)=\frac{1}{2}\left[-a_1(y_1-y_1^0)-a_1(y_2-y_2^0)-a_3(x_1-x_1^0)+a_2(x_2-x_2^0)\right];
$$
$$
W^{\prime}(z_2)=\frac{1}{2}\left[a_1(y_1-y_1^0)+a_1(y_2-y_2^0)-a_2(x_1-x_1^0)+a_3(x_2-x_2^0)\right];
$$

Denote $x_1-x_1^0=\tilde x_1$, $x_2-x_2^0=\tilde x_2$,\ldots the 
deviations from the stationary points. Then we obtain  the following systems
\newpage
a) for the heat-flow equation

\begin{equation}
\begin{cases}
\displaystyle{\frac{d\tilde x_1}{dt}}=a_3\tilde x_1-a_2\tilde x_2,\cr\cr
\displaystyle{\frac {d\tilde x_2}{dt}}=-a_2\tilde x_1+a_3\tilde x_2, \cr\cr
\displaystyle{\frac {d\tilde y_1}{dt}}=a_1\tilde y_1+a_1\tilde y_2, \cr\cr
\displaystyle{\frac{d\tilde y_2}{dt}}=a_1\tilde y_1+a_1\tilde y_2,
\end{cases} 
\label{56}
\end{equation}

b) for the Schr\"{o}dinger equation

\begin{equation}
\begin{cases}
\displaystyle{\frac{d\tilde x_1}{dt}}=-a_1\tilde y_1-a_1\tilde y_2, \cr\cr
\displaystyle{\frac {d\tilde x_2}{dt}}=a_1\tilde y_1+a_1\tilde y_2, \cr\cr
\displaystyle{\frac {d\tilde y_1}{dt}}=a_3\tilde x_1-a_2\tilde x_2, \cr\cr
\displaystyle{\frac{d\tilde y_2}{dt}}=a_2\tilde x_1-a_3\tilde x_2.
\end{cases} 
\label{55}
\end{equation}

Notice that $e_3<e_2\leq 0<e_1$  (\cite{AS}), hence,  $a_1>0$, $a_2>0$ and
the sign of $a_3$ depends on the parameters of the ring. We obtain the systems
of ODEs with the constant coefficients. 

Let the initial positions of the particles be at the points $(\delta_i,\varepsilon_i)$
($i=1,2$), where $\delta_i,\varepsilon_i$ are small parameters. Then  we can obtain an
analitic form for the low of {\it
  small motions} near the stationary points.

\subsection {Heat-flow equation}

The system (\ref{56}) splits in two independent systems for $x_1, x_2$ and
$y_1, y_2$ and we have
$$
\begin{cases}
x_1(t)=C_1 e^{(a_3-a_2)t}+C_2 e^{(a_3+a_2)t},\cr
x_2(t)=C_1 e^{(a_3-a_2)t}-C_2 e^{(a_3+a_2)t}
\end{cases}
$$
with 
$$
\begin{cases}
C_1={\displaystyle{\frac{\delta_1+\delta_2}{2}}}, \cr
C_2={\displaystyle{\frac{\delta_1-\delta_2}{2}}}.
\end{cases}
$$

Since 
$$
a_3+a_2=\displaystyle{\frac{4}{R_1 R_2}\left(e_2-e_3\right)}>0, 
$$
$$
a_3-a_2=\displaystyle{\frac{2}{R_1 R_2}\left(2e_2-\frac{4\eta}{\pi}\right)}<0, 
$$
we have $e^{(a_3+a_2)t}\to\infty$ and $e^{(a_3-a_2)t}\to 0$ as $t\to\infty$. If the 
initial positions of the particles are symmetric with respect to the stationary
points, $\delta_1=-\delta_2=\delta$, we obtain  $x_1(t)=-x_2(t)=\delta e^{(a_3+a_2)t}\to\infty$,
i.e. 
particles move away from the stationary points. For antisymmetric initial
positions $\delta_1=\delta_2=\delta$, and 
$x_1(t)=x_2(t)=\delta e^{(a_3-a_2)t}\to0$. This means that the particles described by the {\it
  linearized} equation are attracted to the stationary
points.

From the system for $y_i$ we have:
$$
\begin{cases}
y_1(t)=\varepsilon_2 e^{2a_1t},\cr
y_2(t)=\varepsilon_2 e^{2a_1t}-(\varepsilon_2-\varepsilon_1).
\end{cases}
$$

Both the solutions are divergent. If at the initial moment the particles are  
at the $x$-axis ($\varepsilon_1=\varepsilon_2=0$) they move along this axis. 

More detail numerical investigation of the motion of the pairs 
``vortex-antivortex'' (not only for $N=2$ but for any even $N$) shows that 
all the
particles move {\it from the stationary points} to the nearest boundary or to
the nearest neighbours and
annihilate. For example, 
initial positions are far enough (the vortices are distributed
uniformly on the same circumference), then they ``don't fill''  each other,  
move to the boundary and annihilate with their reflections (see 
Fig.6a). 

If they are close enough, they attract and annihilate with the nearest
neighbour (see
Fig.6b). Such a  behaviour is similar to the motion of the particles in the
circle and corresponds to the usual physical picture: the
charges with the different signs attract.

\begin{figure}[h]
\unitlength1cm
\begin{minipage}[t]{8.0cm}
\includegraphics[width=7.0cm]{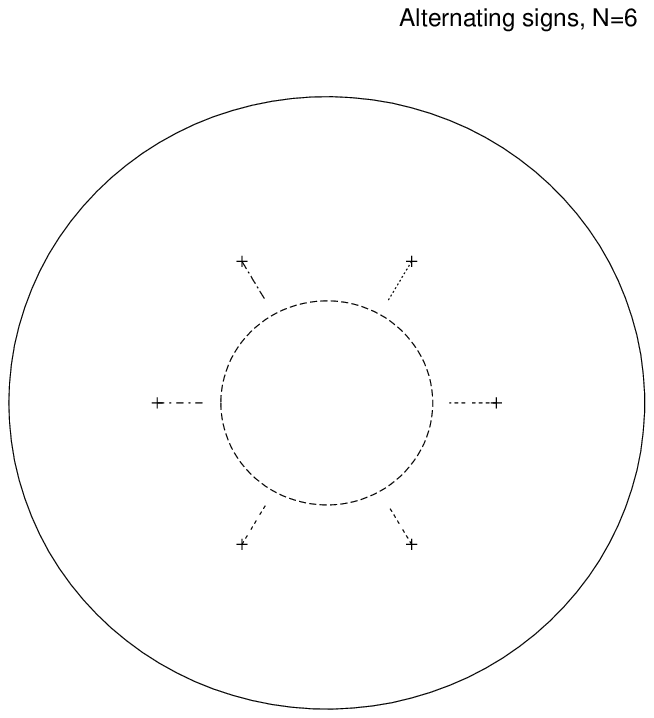}
\begin{center} Fig.6a. Even number of the alternative-sign vortices
\end{center}

\label{fig6a}
\end{minipage}
\hskip 0.5cm
\begin{minipage}[t]{8.0cm}
\includegraphics[width=7.0cm]{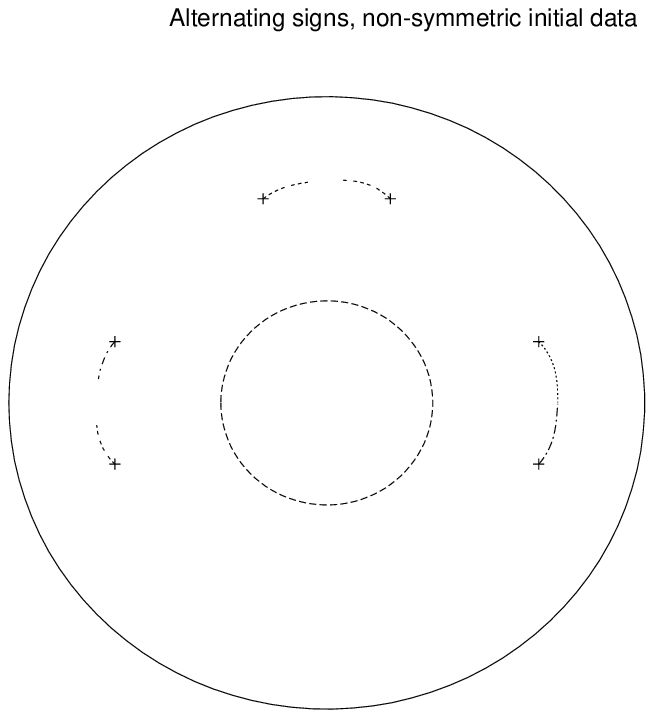}
\begin{center} Fig.6b. Particles are close enough.
\end{center}
\label{fig6b}
\end{minipage}
\end{figure}

If we have all the particles with the equal signs they always move to the
{\it outer} boundary and annihilate. The pictire of motion is the same as for
circle (see Fig. 5b).

\newpage
\subsection{Schr\"{o}dinger equation} 

In the case of the Schr\"{o}dinger equation the system does not split and we have more
complicated motion. We see that 
$\tilde x_2=\tilde x_1+C$ and the second part of the system takes a form
$$ 
\frac{\partial{\tilde y_1}}{\partial t}=a_3\tilde x_1-a_2(C-\tilde x_1), 
$$
$$
\frac{\partial{\tilde y_2}}{\partial t}=a_2\tilde x_1-a_3(C-\tilde x_1).
$$

Eliminating $\tilde y_1$, $\tilde y_2$ from the equation for $\tilde x_1$ we obtain
$$
\frac{d^2\tilde x_1}{dt^2}=-2a_1(a_2+a_3)\tilde x_1-Ca_1(a_3+a_2).
$$
In the standard way we find the solution of the system satisfying the
initial data  $\tilde x_1(0)=\delta_1$, $\tilde x_2(0)=\delta_2$, $\tilde y_1(0)=\varepsilon_1$,
and $\tilde y_2(0)=\varepsilon_2$:
$$
\begin{cases}
\tilde x_1(t)=\displaystyle{\frac{\delta_1-\delta_2}{2}\cos kt-d_2 \sin
kt+\frac{\delta_1+\delta_2}{2}},\cr
\tilde x_2(t)=-\displaystyle{\frac{\delta_1-\delta_2}{2}\cos kt-d_2\sin
kt+\frac{\delta_1+\delta_2}{2}}.
\end{cases}
$$

Here $k$ equals $k=\sqrt{2a_1(a_2+a_3)}$ and $d_2$ is a constant that will
  be determined below.  

The system for $y_i$ with initialy conditions $y_1(0)=\varepsilon_1$ and $y_2(0)=\varepsilon_2$ yields 
\begin{equation}
\begin{cases}
\tilde y_1(t)=\displaystyle{\frac{(\delta_1-\delta_2)(a_2+a_3)}{2k}\sin kt+\frac{\varepsilon_1+\varepsilon_2}{2}\cos
kt+\frac{(\delta_1+\delta_2)(a_3-a_2)}{2}t+\frac{\varepsilon_1-\varepsilon_2}{2}},\cr\cr
\tilde y_2(t)=\displaystyle{\frac{(\delta_1-\delta_2)(a_2+a_3)}{2k}\sin kt+\frac{\varepsilon_1+\varepsilon_2}{2}\cos
kt+\frac{(\delta_1+\delta_2)(a_2-a_3)}{2}t-\frac{\varepsilon_1-\varepsilon_2}{2}}.
\label{y12}
\end{cases}
\end{equation}

We obtain the constant $d_2$ from the equation for $\tilde x_2$: 
$\dot x_2(0)=a_1(\varepsilon_1+\varepsilon_2)$. Namely, 
$$
d_2=-\frac{a_1(\varepsilon_1+\varepsilon_2)}{k}.
$$

We have {\it periodic solutions} for all the variables if $\delta_1=-\delta_2=\delta$ and
$\varepsilon_1=\varepsilon_2=0$, i.e. if the initial positions of the vortices are symmetric
with respect to y-axis. Thus 
$$
\begin{cases}
\tilde x_1=\delta\cos kt,  \hskip 2.3cm \tilde y_1=\displaystyle{\frac{\delta(a_2+a_3)}{k}\sin kt},
\cr
\tilde x_2=-\delta\cos kt, \hskip 2cm \tilde y_2=\displaystyle{\frac{\delta(a_2+a_3)}{k}\sin kt}.
\end{cases}
$$

The vortices describe elliptic trajectories.

In general any number of the pairs
``vortex-antivortex'' with symmetric initial positions moves along the closed trajectories (see
Fig.7a for $N=2$ and Fig.7b for $N=6$). If the initial positions are close to the stationary points, the
trajectories are "narrower" and "shorter".

\bigskip
\begin{figure}[h]
\unitlength1cm
\begin{minipage}[t]{8.0cm}
\includegraphics[width=6.0cm]{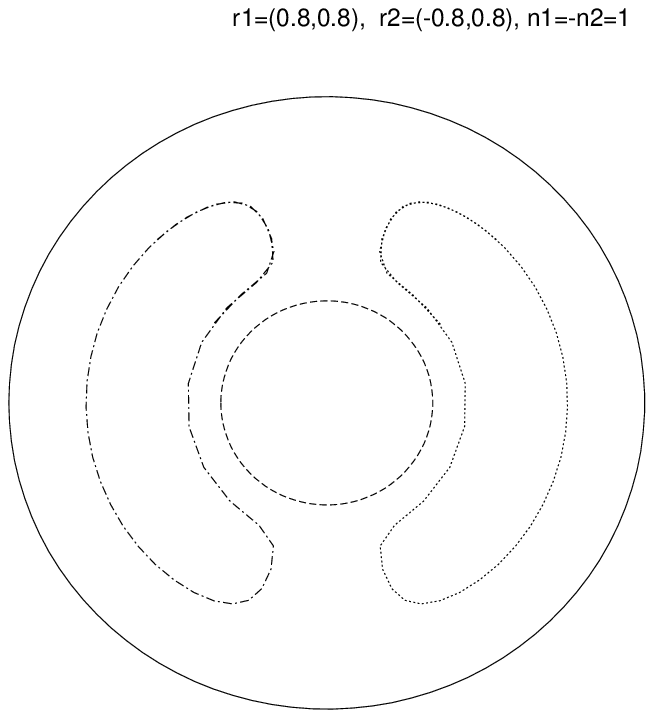}

\begin{center} Fig.7a. 
\end{center}

\label{fig3}
\end{minipage}
\hskip 0.5cm
\begin{minipage}[t]{8.0cm}
\includegraphics[width=6.0cm]{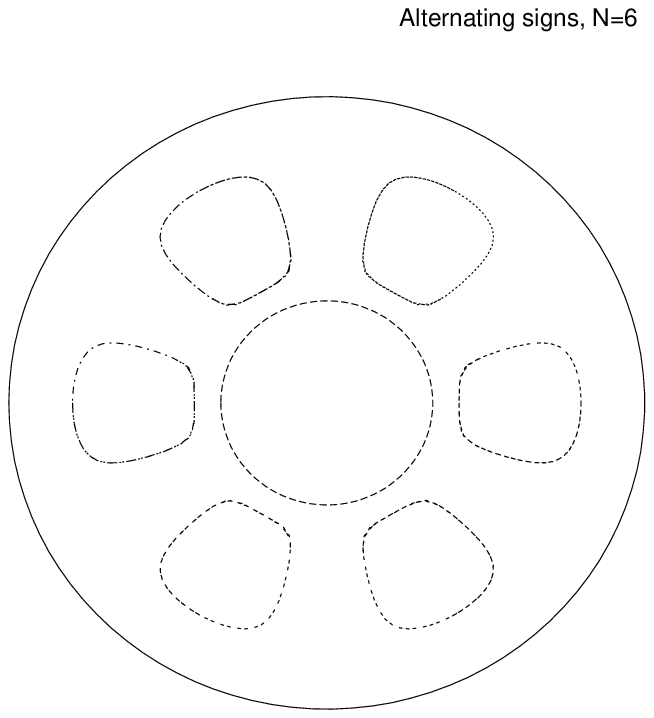}
\begin{center} Fig.7b.
\end{center}
\label{fig4}
\end{minipage}
\end{figure}

Non-symmetric initial positions lead  to a very complicated motion due to the
terms which are linear in $t$ (see (\ref{y12})). The
velocities of the vortices are different and the picture of the motion for the
linear problem 
strongly  depends 
on the initial data.

Thus, if we have the particles with the different signs with arbitrary initial
positions  we obtain  chaotic non-periodic motion
(see Fig.8a for $N=5$). Even the initial positions are symmetric but the
number of particles is odd, the motion is chaotic (Fig.8b).

\bigskip
\begin{figure}[h]
\unitlength1cm
\begin{minipage}[t]{8.0cm}
\includegraphics[width=6.0cm]{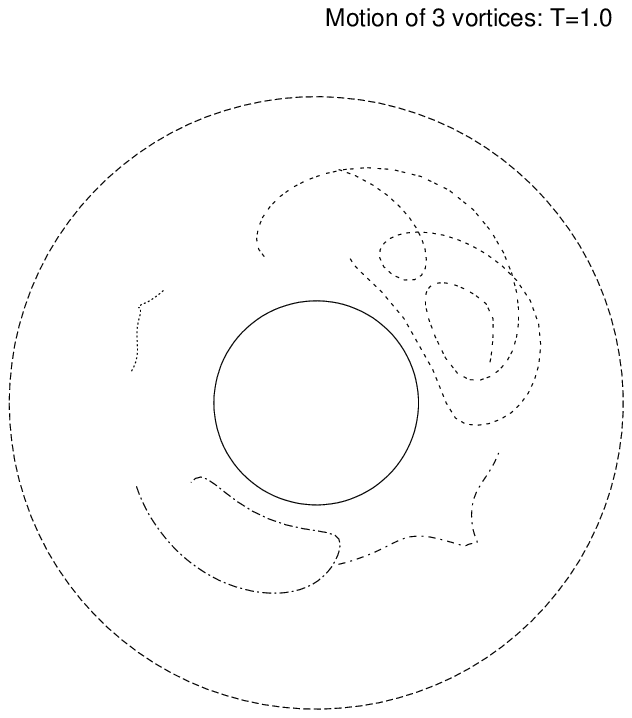}
\begin{center} Fig.8a. 
\end{center}

\label{fig9}
\end{minipage}
\hskip 0.5cm
\begin{minipage}[t]{8.0cm}
\includegraphics[width=6.0cm]{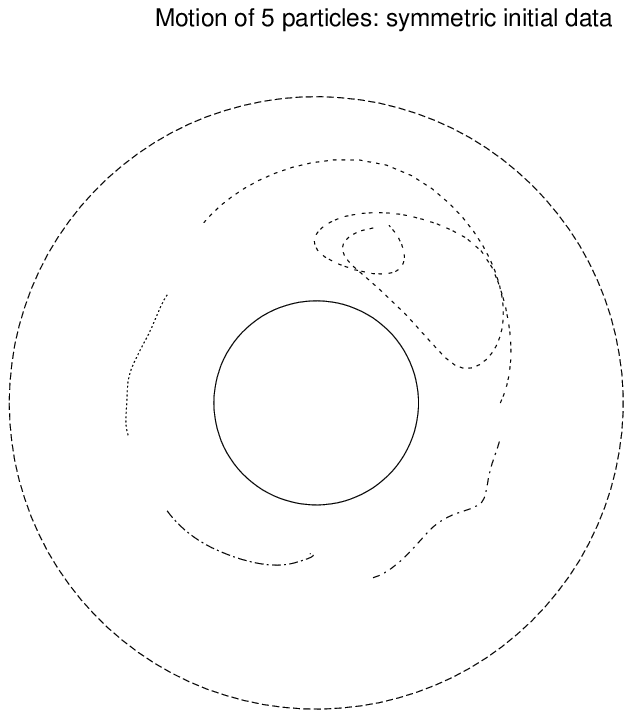}
\begin{center} Fig.8b. 
\end{center}
\label{fig10}
\end{minipage}
\end{figure}

\vskip 1cm
{\bf Motion "vortex-vortex"}. The stationary points in this case is not so convenient to calculate the
coefficients of the linear system. But it is obvious that if the initial
positions are symmetric the vortices move by the same way. It means that the particles move independently in the
same direction on the same circumference as it was for the circle
(see. Fig.2b). If the initial positions
are not symmetric, the vortices move globally in the same directions but
along the comlicated trajectories with the loops (see, for example,  Fig. 9 for $N=3$).  

\bigskip
\begin{figure}[t]
\unitlength1cm
\begin{minipage}[t]{8.0cm}
\includegraphics[width=6.0cm]{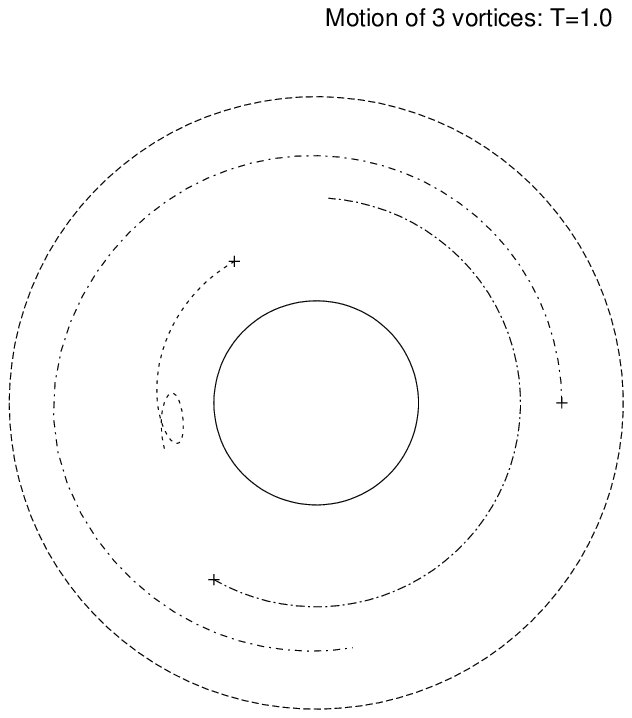}
\begin{center} Fig.9a. 
\end{center}

\label{fig9}
\end{minipage}
\hskip 0.5cm
\begin{minipage}[t]{8.0cm}
\includegraphics[width=6.0cm]{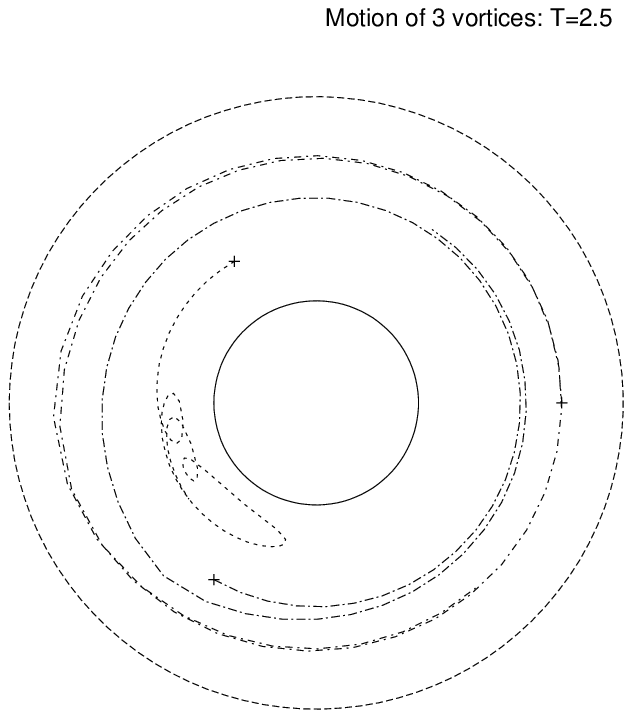}
\begin{center} Fig.9b. 
\end{center}
\label{fig10}
\end{minipage}
\end{figure}

So, for Schr\"{o}dinger equation we have very different pictures of motion
depend on the initial positions of the vortices.

\section*{Acknowledgements}

This work is supported by the NATO scientific exchange grant (2002). The author
is grateful to the Laboratory Jacques-Louis Lions at the University Paris-6
for kind hospitality. 

The author expresses sincere gratitude to Prof. F.Bethuel for the  interest to
this work and many useful discussions. Also the help  of Dr. D.Smets is 
greatly 
appreciated.

{}

\end{document}